\documentclass[aip,reprint]{revtex4-1}

\usepackage[version=3]{mhchem} 
\usepackage{xcolor, float, tabularx, graphicx, booktabs}

\AtBeginDocument{%
  \heavyrulewidth=.08em
  \lightrulewidth=.05em
  \cmidrulewidth=.03em
  \belowrulesep=.65ex
  \belowbottomsep=0pt
  \aboverulesep=.4ex
  \abovetopsep=0pt
  \cmidrulesep=\doublerulesep
  \cmidrulekern=.5em
  \defaultaddspace=.5em
}

\begin{document}

\author{Maximilian R. Becker}
\affiliation{Theoretical Bio- and Soft Matter Physics, Freie Universität Berlin, Arnimallee 14, Berlin 14195, Germany}
\author{Ruben Cruz}
\affiliation{Experimental Molecular Biophysics, Freie Universität Berlin, Arnimallee 14, Berlin 14195, Germany}
\author{Kenichi Ataka}
\affiliation{Experimental Molecular Biophysics, Freie Universität Berlin, Arnimallee 14, Berlin 14195, Germany}
\author{Joachim Heberle}
\affiliation{Experimental Molecular Biophysics, Freie Universität Berlin, Arnimallee 14, Berlin 14195, Germany}
\author{Roland R. Netz}
\affiliation{Theoretical Bio- and Soft Matter Physics, Freie Universität Berlin, Arnimallee 14, Berlin 14195, Germany}
\email{}

\title{Microscopic Structure and Dynamics of Interfacial Water at Fluorinated vs Nonfluorinated Surfaces - Insights from Ab-Initio Simulations and IR Spectroscopy}





\begin{abstract}

Per- and polyfluoroalkyl substances are a class of synthetic chemical compounds widely used as coatings to lower surface energies. Yet the microscopic mechanisms of their weak interaction with water and organic compounds remain poorly understood. Here, we perform large-scale density-functional-theory molecular dynamics simulations to investigate water at self-assembled monolayers (SAMs) of fluorinated and non-fluorinated hydrocarbons. We analyze the interfacial water structure and compare it to the prototypical hydrophobic air–water interface. The interfacial water structure at both SAMs closely resembles that at the air–water interface, featuring a distinct depletion layer and a two-dimensional hydrogen-bond network parallel to the surface. Computed anisotropic infrared spectra reproduce key experimental signatures observed in surface-enhanced infrared absorption spectroscopy (SEIRAS), including the presence of free OH vibrations directly probing the local surface–water interactions. Notably, while the free OH stretch at the hydrocarbon SAM–water interface exhibits a red shift relative to the air–water interface, indicative of weak binding, the fluorinated SAM–water interface displays a weakly blue-shifted free OH mode, in agreement with experiment. This frequency behavior defies common interpretations based on the vibrational Stark effect. Further, we show that the reorientation dynamics of water molecules are significantly slower near the fluorinated surface, which we correlate with spectral line shapes, an effect rather expected at hydrophilic surfaces. This indicates that fluorinated SAMs, despite being macroscopically more hydrophobic than their unfluorinated counterparts, exhibit spectroscopic characteristics that neither qualify it as hydrophobic nor hydrophilic.

\end{abstract}

    \maketitle

\section{Introduction}

Per- and polyfluoroalkyl substances (PFAS) are a class of synthetic organic chemical compounds characterized by alkyl chains with multiple fluorine atoms attached. Due to their ability to modify the physical and chemical properties of various surfaces, these compounds hold significant scientific and technological importance in fields as diverse as nanoelectronics \cite{klauk2010organic, xiang2016molecular}, biomolecular science \cite{cametti2012fluorous, lv2021fluoropolymers}, and electrochemistry \cite{ozden2020high}. One of the most intriguing characteristics of PFAS is that the replacement of hydrogen atoms with fluorine atoms typically leads to an increase in hydrophobicity, as well as enhanced lipophobicity, a phenomenon known as omniphobicity. In the context of surface science, this means that PFAS-based surface coatings exhibit very low surface energies. Consequently, the incorporation of fluorochemicals is one of the primary techniques for achieving lower wettability of surfaces \cite{tuteja2007designing, si2015superhydrophobic, parvate2020superhydrophobic}. Other characteristics that make PFAS-based coatings attractive include their high thermal and chemical stability \cite{tuteja2007designing, cametti2012fluorous, soto2018short}. However, it has been found that the widespread use and long natural lifetimes of PFAS have led to their accumulation in the environment, and PFAS contamination has been linked to a variety of environmental problems and human diseases \cite{brunn2023pfas, jian2018short, fenton2021per}. Thus, there is an ongoing effort to develop alternatives to PFAS-based coatings with comparably low surface energies. Understanding the microscopic mechanisms that govern the physicochemical characteristics of PFAS-based surface coatings can guide this effort. 

Self-assembled monolayers of fluorinated alkanethiols (FSAMs) adsorbed on various metal surfaces provide a widely used platform to study fluorous thin films under controlled environments. Compared to their unfluorinated counterparts (alkanethiol self-assembled monolayers, HSAMs), they show larger contact angles for a variety of different liquids \cite{graupe1999oriented, colorado2003wettabilities, barriet2003fluorinated} as well as lower surface-liquid friction coefficients \cite{kim1997systematic, li2019molecular, li2021fluorination}. Molecular modeling has proven to be a valuable tool for understanding PFAS properties. For individual fluorinated hydrocarbons in aqueous bulk solution, it has been established that the main cause of their increased hydrophobicity, compared to their non-fluorinated counterparts, is the amount of work required to form a larger cavity to encompass fluoroalkyls. This energetic contribution dominates over the increased electrostatic binding affinity between water molecules and the polar C-F bonds, and the differences in dispersive interactions \cite{dalvi2010molecular}. In previous work, a similar mechanism has been identified that accounts for the increased hydrophobicity of FSAMs compared to HSAMs: Molecular dynamics (MD) simulations based on empirical force fields (FFs) have shown that the dominant interaction forces between nonpolar SAMs and water are dispersive interactions, which are approximately proportional to the grafting density of the SAMs. Due to the inherent size of fluorine atoms and the helicity of perfluorinated linear hydrocarbons, the grafting density of FSAMs is considerably lower than that of HSAMs, resulting in their strongly hydrophobic behavior \cite{dalvi2010molecular, carlson2021hydrophobicity}. At first glance, this result may seem rather counterintuitive. It is well established that fluorinated and non-fluorinated hydrocarbons have very different charge distributions which leads to differences in the hydration structure of single molecules in water \cite{robalo2019hydrophobic}. Yet, according to the reported FF-MD simulations this has only little affect on the hydrophobicity of fluorinated SAMs \cite{dalvi2010molecular, carlson2021hydrophobicity}. 

However, the available theoretical studies regarding the interaction of water with fluorinated and non-fluorinated SAMs mostly rely on empirical nonpolarizable force fields designed to reproduce certain thermodynamic properties of bulk liquids. It remains unclear whether such force fields can accurately describe interactions of water at interfaces, where the hydrogen bond environment deviates significantly from the bulk case. For instance, it is known that water molecules change their dipole moment from 1.85 D in vacuum \cite{Clough1973} to 2.9 D \cite{Badyal2000} in the liquid phase, an effect that is clearly missing in non-polarizable force field models. Thus, simulations at a higher level of molecular modeling are necessary for an accurate understanding of water at fluorinated surfaces.

Experimentally, the microscopic structure of water at interfaces has been probed using various spectroscopic techniques, including IR and THz absorption spectroscopy \cite{LeCaer2011, Knight2019, pezzotti2021molecular, Ruiz-Barragan2022, zhang2025infrared}, and surface-specific nonlinear spectroscopic methods, such as sum frequency generation (SFG) \cite{Du1993, du1994surface, perakis2016vibrational, gonella2021water} . Different spectroscopic fingerprints of water at hydrophobic interfaces have been established. Particular well studied is the emergence of so-called 'dangling' or 'free' OH bonds \cite{Du1993, du1994surface, perakis2016vibrational, pezzotti2021molecular}, corresponding to water molecules in the outermost layer, whose OH bonds point towards the hydrophobic interface and are therefore not hydrogen bonded. These water molecules give rise to distinct spectroscopic markers in SFG spectra: the emergence of a narrow band at around 3700~cm$^{-1}$, a frequency higher than the typical bulk OH stretch peak. Signatures of the same population of water molecules have also been analyzed using x-ray absorption spectroscopy \cite{velasco2014structure}  and surface enhanced infrared absorption spectroscopy \cite{ataka1996potential}. Since spectroscopic signatures of free OH groups are typically found at hydrophobic interfaces, the intensity of the corresponding band has been proposed as a microscopic marker of hydrophobicity \cite{tang2020molecular}. However, recent experiments have shown that free OH groups can also be found at macroscopically hydrophilic surfaces \cite{tuladhar2017insights, cyran2019molecular, wang2019gate}. Instead, the vibrational frequency of free OH groups measured in sum frequency generation spectroscopy has been proposed as an indicator of hydrophobicity \cite{pezzotti2021molecular}. Comparing SFG spectra of water at a range of different surfaces, it was shown that the spectral peak corresponding to the vibration of free OH groups shifts to lower frequencies at more hydrophilic interfaces. This can be interpreted within the framework of the vibrational Stark effect: Stronger electrostatic interactions between water molecules and surfaces, which give rise to the increased hydrophilicity of the surface, lead to a red-shift of the free OH vibrational feature \cite{auer2008ir, bishop1993vibrational}. A different framework for interpreting vibrational frequency shifts was recently introduced by Br\"unig et al, in which vibrational blue shifts were related to frequency-dependent friction \cite{bruenig2022time}.

In this work, we report on large-scale density-functional-theory-based (DFT) MD simulations of the SAM-water interface for  perfluoroalkylthiol-SAMs and alkanethiol-SAMs. We additionally compare our simulation results to the prototypical hydrophobic interface, the air-water interface. Our findings demonstrate that the microscopic water structure predicted by DFT-MD aligns well with results from empirical force field simulations. Moreover, key microscopic descriptors of the surface's hydrophobicity, such as the thickness of the interfacial depletion layer, show qualitative consistency with both experimental data and previous computational predictions. The water's hydrogen bond network in the vicinity of all three interfaces is remarkably similar: we observe the formation of a layer of water molecules that are predominantly hydrogen bonded with other water molecules in the same layer, as was previously reported for various hydrophobic surfaces \cite{pezzotti20172d}. 

To verify our results, we calculate the anisotropic infrared absorption spectrum of the interfacial water and compare it to surface enhanced infrared absorption spectroscopy (SEIRAS) experiments. Indeed, we observe the spectroscopic signatures of free OH groups in our SEIRAS experiments as well as our DFT-MD simulations. As these bonds are in direct contact with the SAMs, their vibrational properties reveal important information about surface-water interactions. Interestingly, the correlation between free OH vibrational frequency and macroscopic surface hydrophobicity does not hold for the SAM-water interfaces reported in this work. While we observe a red shift of the free OH peak for the HSAM-water interface relative to the air-water interface, we observe a weak blue shift at FSAM-water interface. Further analysis of the vibrations of free OH groups indeed show that the FSAM-water interface defies interpretation in terms of the vibrational Stark effect: where such an interpretation would suggest that OH groups that are in close vicinity to negatively charged fluorine atoms would vibrate at a lower frequency due to a softening of the OH bond, we observe the opposite - the closer an OH group is to fluorine atoms, the higher is its vibrational frequency. The SEIRAS experiments confirm our theoretical predictions that the free OH peak is indeed observed at a higher frequency at the FSAM-water interface than at the HSAM-water interface. These findings support previous theoretical results, indicating that dispersive interactions, rather than electrostatic interactions, dominate the FSAM-water interface \cite{dalvi2010molecular, carlson2021hydrophobicity}.

In addition to their frequency, we analyze the vibrational line shapes of the free OH spectral features, which we interpret in terms of vibrational lifetimes of free OH groups. We find significantly different line widths of the free OH peaks, with the narrowest peak observed at the FSAM-water interface and the broadest peak at the air-water interface. This reflects that water molecules are less mobile in the interfacial layer at SAMs in general, and at FSAMs in particular. Our analysis reveals that the reduction in line widths is primarily caused by a strong slow-down in the reorientation of water molecules in the vicinity of fluorinated surfaces, compared to both the non-fluorinated interface and the air-water interface. While the deceleration of reorientational dynamics can be observed in the vicinity of either very hydrophobic or very hydrophilic surfaces, an effect as strong as the one observed at the FSAM water interface, would rather be associated with hydrophilic surfaces \cite{romero2009effect, stirnemann2011non,gekle2012anisotropy}.

\section{Results and discussion}

We conduct DFT-MD simulations of the HSAM-water interface (Fig.~\ref{fig:fig_1}\,B), comprising 16 n-octane molecules (Fig.~\ref{fig:fig_1}\,D) and 198 water molecules as well as of the FSAM-water interface (Fig.~\ref{fig:fig_1}\,C) comprising 16 semifluorinated molecules CH$_3$CH$_2$(CF$_2$)$_3$CF$_3$ (Fig.~\ref{fig:fig_1}\,D) and 286 water molecules. Carbon atoms at the bottom of the SAMs are restrained to a hexagonal grid (Fig.~\ref{fig:fig_1}\,D-E) using harmonic restraint potentials. The lattice constants are chosen to align with the experimentally determined values of a = 4.97~$\mathrm{\AA}$ for HSAMs and 5.9~$\mathrm{\AA}$ for FSAMs \cite{chidsey1990chemical,alves1993atomic, barriet2003fluorinated}. We compare our DFT-MD simulations to FF-MD simulations comprising 200 nonfluorinated and semifluorinated hexane molecules each, and 2089 and 2928 water molecules, respectively. Water molecules are modeled using the SPC/E model \cite{berendsen1987missing}, while hydrocarbons are modeled using the OPLS/AA FF \cite{jorgensen1996development} and its extension for perfluorinated hydrocarbons \cite{watkins2001perfluoroalkanes}. In a previous work, we have adapted this FF to contain parameters for semifluorinated hydrocarbons \cite{carlson2021hydrophobicity}. In the same work, we have shown that the employed FFs can quantitatively  reproduce the experimental contact angles of water droplets on both HSAMs and FSAMs with high accuracy. The primary mechanism responsible for the increased hydrophobicity of FSAMs, relative to HSAMs, was identified as the larger lattice spacing of semifluorinated monolayers \cite{carlson2021hydrophobicity}. Further simulation details can be found in the Methods section and snapshots of the full simulation setups are shown in Sec.~S1 of the supplementary material. 

The prototypical hydrophobic interface is the air-water interface. To identify surface-specificity of our results, we compare to a previously published\cite{becker2022electrokinetic} DFT-MD simulation of 352 water molecules within a periodic box of size $2\times2\times6$~nm$^3$. These simulations were conducted using the same exchange-correlation functional and simulation parameters, rendering them comparable to the SAM-water interface simulations reported in this work. The same system was also simulated using FF-MD simulations employing the SPC/E force field \cite{berendsen1987missing}. All details of the air-water simulations are provided in \cite{becker2022electrokinetic}.

\subsection{Structure of interfacial water}

\begin{figure*}[t]
    \centering
    \includegraphics[width=0.85\linewidth]{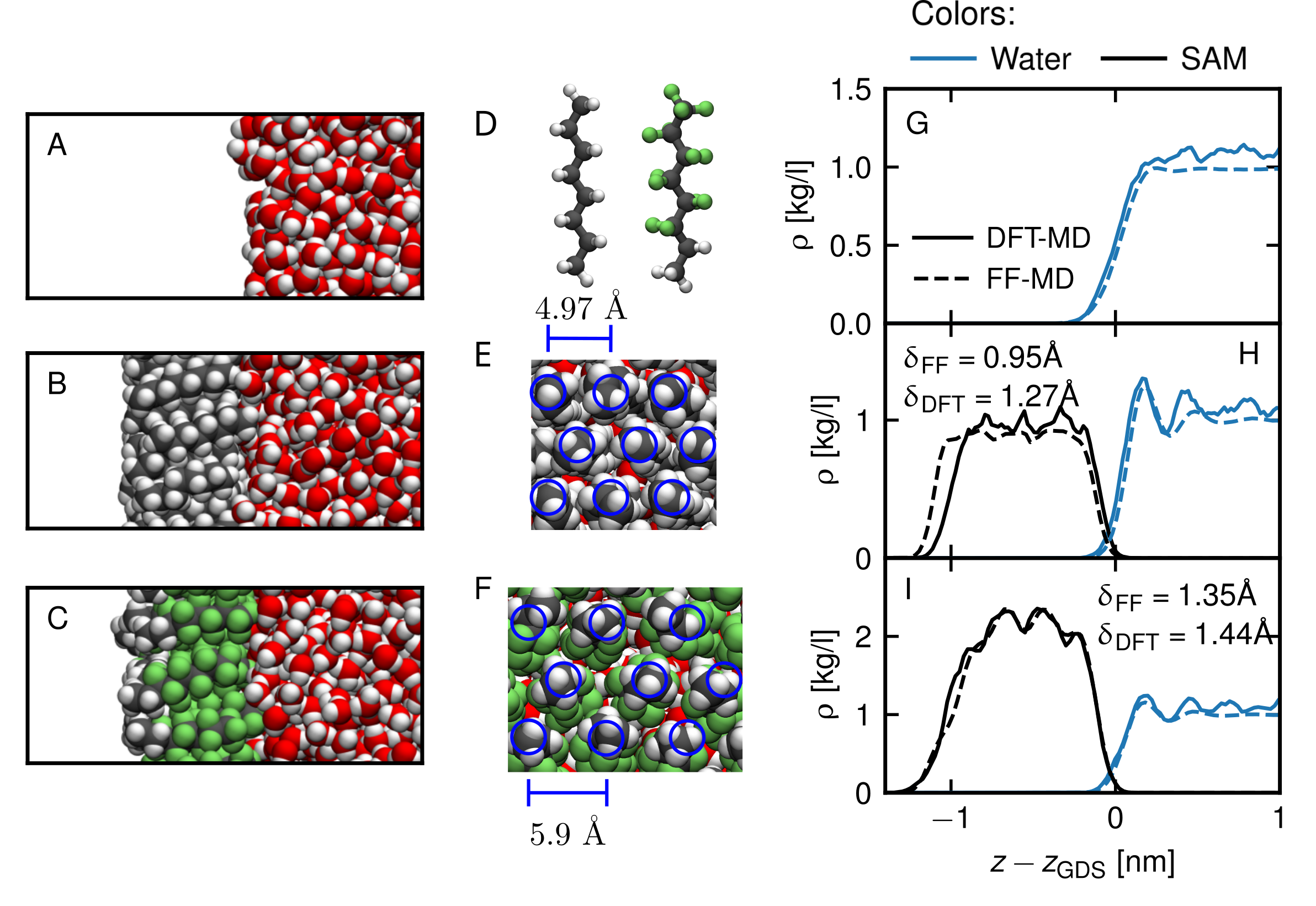}
    \caption{\textbf{A}-\textbf{C}: Simulation snapshots from DFT-MD simulations of the three model interfaces: air-water, HSAM-water containing n-hexane molecules (\textbf{D}, left), and FSAM-water containing the semifluorinated compound CH$_3$CH$_2$(CF$_2$)$_3$CF$_3$ (\textbf{D}, right). To keep the SAMs in place, carbon atoms at the bottom of the SAMs are restricted to a hexagonal close lattice indicated in panels \textbf{E} and \textbf{F} by blue circles. \textbf{G}-\textbf{I}: Density profiles in the vicinity of all three interfaces as a function of the distance to the Gibb's dividing surface of the water calculated according to Eq.~\ref{eq:theory:gibbs_dividing}. The thicknesses of the depletion layers $\delta$, defined in Eq.~\ref{eq:sams:depletion_layer}, are given in the legends.}
    \label{fig:fig_1}
\end{figure*}

Experimentally, a surface's hydrophobicity is typically characterized by the contact angle that a water droplet forms on it. While the accurate determination of contact angles via FF-MD simulations is feasible, it requires either the simulation of very large systems when explicit water droplets are considered \cite{carlson2021hydrophobicity,de1999dynamic,kanduvc2017going} or very long simulation times when surface energies are estimated from the anisotropy of the pressure tensor \cite{carlson2024modeling}. Both methods are prohibitively expensive for implementation with DFT-MD simulations. Instead, to establish the hydrophobicity of HSAMs and FSAMs through DFT-MD simulations, we rely on comparing microscopic observables that characterize surface energy between our DFT-MD simulations and FF-MD simulations for which contact angles have previously been established \cite{carlson2021hydrophobicity}. For instance it has been shown that the density depletion between a solid surface and water is strongly correlated to the contact angle \cite{janecek2007interfacial, huang2008water, sendner2009interfacial, carlson2025sub}. Since mass density profiles converge already for short simulation times, this depletion is a good observable to compare the surface energy of different surfaces from DFT-MD simulations. 

The extracted mass density profiles of water in Fig.~\ref{fig:fig_1}\,G-I are notably similar when comparing DFT-MD and FF-MD simulations at all three interfaces, air-water, HSAM-water and FSAM-water. In particular, for the HSAM-water and FSAM-water interfaces, the widths and heights of the first water density peaks close to the SAMs align well, indicating good agreement regarding the interfacial water structure between the two simulation techniques. The density depletion is then quantified by the thickness of a depletion layer, here denoted as $\delta$, which is defined by the separation between the respective Gibbs dividing surfaces (GDS) of the SAMs and the water. Using 
\begin{align}
    z_\mathrm{GDS}^\mathrm{l} = z_0 + \int\limits_{z_0}^{z_\mathrm{l}}  \mathrm{d}z \, \left[1  - \frac{\rho_\mathrm{l}(z)}{\rho_\mathrm{l}^\mathrm{bulk}}\right]\,,
    \label{eq:theory:gibbs_dividing}
\end{align}
to determine the Gibbs's dividing surface position of the liquid, $\delta$ can be calculated according to \cite{mamatkulov2004water}
\begin{align}
    \delta = z_\mathrm{GDS}^\mathrm{s} - z_\mathrm{GDS}^\mathrm{l} = \int\limits_\mathrm{z_\mathrm{s}}^{z_\mathrm{l}}\mathrm{d}z \left(1 - \frac{\rho_\mathrm{s}(z)}{\rho_\mathrm{s}^\mathrm{bulk}} - \frac{\rho_\mathrm{l}(z)}{\rho_\mathrm{l}^\mathrm{bulk}}\right)\, , \label{eq:sams:depletion_layer}
\end{align}
where $\rho_\mathrm{s}(z)$ and $\rho_\mathrm{l}(z)$ refer to the z-dependent mass densities of the surface and the liquid and $z_\mathrm{s}$, $z_\mathrm{l}$ are positions well inside the surface and the liquid bulk, where the respective densities are given by $\rho_\mathrm{s}^\mathrm{bulk}$ and $\rho_\mathrm{s}^\mathrm{bulk}$. Note that defining the bulk density of the SAMs from MD density profiles is not straightforward. While, with sufficient sampling, the density profile of water forms a plateau far from the interface (Fig.~\ref{fig:fig_1}\,G-I, blue lines) which defines the bulk density, the mass density of the SAMs oscillates, even for extended sampling times, due to its high spatial ordering. In Sec.~S2 of the supplementary material, we discuss how to address this issue and show that a unique value for $\delta$ can be determined. 

The calculated thicknesses of the depletion layers, which are given in Fig.~\ref{fig:fig_1}\,H-I, agree qualitatively between DFT-MD and FF-MD: both methods exhibit a larger depletion at the FSAM-water interface compared to the HSAM-water interface, indicating increased hydrophobicity of the FSAM, which aligns with previous experimental contact angle measurements \cite{chidsey1990chemical, carlson2021hydrophobicity}. It is noteworthy that the FF-MD setup employed in this work has previously been found to accurately predict contact angles of HSAMs and FSAMs \cite{carlson2021hydrophobicity}.

\begin{figure*}[t]
    \centering
    \includegraphics[width=0.85\linewidth]{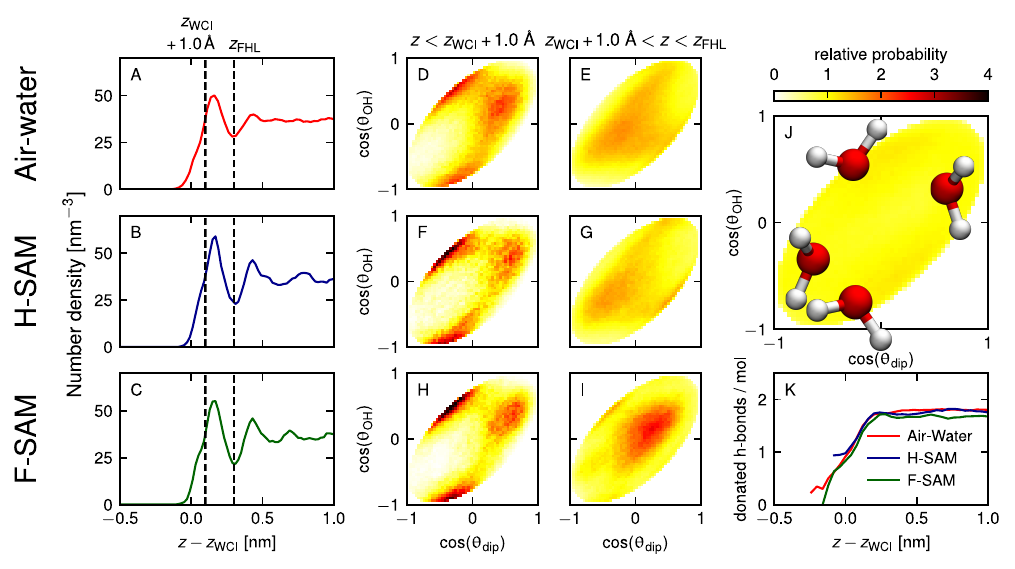}
    \caption{Interfacial water structure at different interfaces from DFT-MD simulations. For comparison, FF-MD results are shown in Sec.~S3 of the supplementary material. \textbf{A}-\textbf{C}: Molecular number density profiles of water molecules as a function of the distance from the centers of mass to the WCI. For all three interfaces the first minimum of the density profile is found at the same distance to the WCI, $z_\mathrm{fhl} - z_\mathrm{WCI}=3.0~\mathrm{\AA}$, marked with a vertical dashed line. \textbf{D}-\textbf{I}: Joint probability angular distributions of the orientations of molecular water dipole moments $\theta_\mathrm{dip}$ and OH bonds $\theta_\mathrm{OH}$ normalized to their respective homogeneous distributions (Eq.~\ref{eq:homogeneous_dist} in the  methods section). Data is shown for two distinct slabs, as defined by the positions of the molecules' center of mass, with boundaries indicated by dashed vertical lines in panels A-C. Positive values of $\cos \theta_\mathrm{dip}$ and $\cos \theta_\mathrm{OH}$ indicate that the molecules' dipole or the OH vector points towards the bulk phase, respectively. \textbf{J}: Example orientations of water molecules for various combinations of $(\cos \theta_\mathrm{dip}, \cos \theta_\mathrm{OH})$. \textbf{K}: Profile of the number of donated hydrogen bonds per water molecule, binned with respect to the center of mass.}
    \label{fig:orientations}  
\end{figure*}

After establishing that our DFT-MD simulations reproduce the interfacial properties found in experimental and previous theoretical work, we analyze how this relates to the microscopic water structure in the vicinity of the interface, particularly in comparison with the prototypical hydrophobic interface: the air-water interface. It is important to note that, due to interfacial nano roughness and the occurrence of capillary waves at aqueous interfaces, the profiles of interfacial properties depend on the scale at which they are observed \cite{sedlmeier2009nanoroughness}. To mitigate the influence of varying surface roughness on the comparison between the air-water and SAM-water interfaces, we calculate the Willard-Chandler interface (WCI), which constructs an instantaneous and locally flexible interface with respect to which interfacial properties can be compared \cite{Willard2010}. Indeed, when determined in the laboratory frame, the water density profile at the air-water interface is fairly broad and approximately sigmoidal in shape (Fig.~\ref{fig:fig_1}\,G), while a stratified density is observed for the SAM-water interface (Fig.~\ref{fig:fig_1}\,H-I). In contrast, when determined with respect to the WCI, the molecular number density profiles of all three interfaces reveal remarkably similar stratification (Fig.~\ref{fig:orientations}\,A-C), indicating that the microscopic molecular structure of water at alkane SAMs is primarily governed by water-water interactions rather than water-surface interactions. For instance, the position of the first minimum of the density profile, which in our notation defines the first hydration layer (FHL) of the interface, is found to be identical for all three interfaces at $z_\mathrm{fhl}-z_\mathrm{WCI}=3.0$~$\mathrm{\AA}$. Additionally, the density maxima are of similar magnitude. In Sec.~S3 of the supplementary material we shows that  the same density stratification is found from FF-MD simulations, corroborating our conclusion that DFT-MD and FF-MD simulations predict similar water-surface interactions. 

To study the arrangement of water in the first hydration layer of the interface, we perform orientational analysis of water molecules. The orientation of a water molecule can be characterized by the two angles $\theta_\mathrm{dip}$ and $\theta_\mathrm{OH}$, which correspond to the angles that the molecule's dipole vector and the angle an OH bond form with the interfacial normal, respectively \cite{sedlmeier2008water}. The joint orientational distribution functions in Fig.~\ref{fig:orientations}\,D,F,H, reveal a strongly anisotropic orientation of water molecules in the outermost layer of water ($z < z_\mathrm{WCI}+1.0~\mathrm{\AA}$) at all three interfaces. In this region, consistent with the observations from the density profiles depicted in Figs.~\ref{fig:orientations}\,A-C, the orientation distribution functions further demonstrate that the structure of interfacial water is remarkably similar across all three interfaces considered. Regarding orientation, we find two major populations of water molecules: The first population consists of molecules in which one of the OH bonds points away from the bulk phase toward the interface ($\cos\theta_\mathrm{OH}\approx-1$), while the other OH bond is oriented toward the bulk phase. This population corresponds to the two probability maxima in Figs.~\ref{fig:orientations}D-F around ($\cos\theta_\mathrm{dip}\approx -0.3, \cos\theta_\mathrm{OH}\approx-1$) and ($\cos\theta_\mathrm{dip}\approx -0.3, \cos\theta_\mathrm{OH}\approx0.5$). Corresponding molecular orientations are indicated in Fig.~\ref{fig:orientations}\,J. Note that, due to the fact that each water molecule has two OH bonds, both maxima in the probability distribution correspond to the same population of water molecules. The second population that can be found in the outermost layer of the interface, consists of water molecules with an in-plane orientation, whose OH bonds point only slightly toward the bulk phase (the corresponding maximum is located around ($\cos\theta_\mathrm{dip}\approx 0.5, \cos\theta_\mathrm{OH}\approx0.5$) and the corresponding orientation is shown in Fig.~\ref{fig:orientations}\,J). The orientation of water molecules in the remainder of the first hydration layer in the region $z_\mathrm{WCI}+1.0~\mathrm{\AA} < z < z_\mathrm{fhl}$ is shown in Fig.~\ref{fig:orientations}\,G-I. Overall, we observe less orientational anisotropy compared to the immediate vicinity of the interface. The main feature is an increased tendency for molecules to align their OH bonds parallel to the interface. These results are consistent with previous experimental and theoretical work \cite{sedlmeier2008water, pezzotti20172d, pezzotti20182d, pezzotti2021molecular, lehmann2025beyond} that found that at hydrophobic interfaces, water forms a layer of molecules, which are strongly connected to each other via in-plane hydrogen bonds.  

The outward-pointing OH bonds observed in the outermost interfacial water cannot act as donors for hydrogen bonds, such that the average number of donated hydrogen bonds per water molecule decreases in the vicinity of all three interfaces (Fig.~\ref{fig:orientations}\,K). These free OH groups give rise to characteristic features in vibrational spectra. 

\subsection{Linear IR spectroscopy of interfacial water}

In the vicinity of interfaces, the frequency-dependent electric susceptibility, which is measured in IR absorption spectroscopy, is position-dependent and anisotropic, giving rise to distinct IR spectra recorded for radiation polarized parallel ($\parallel$) and perpendicular ($\perp$) to the interface \cite{becker2024interfacial}. The two different components of the electric susceptibility can be extracted from DFT-MD simulations using the Green-Kubo relations \cite{becker2024interfacial}
\begin{align}
    \chi_\parallel (z, \omega) &= \frac{\phi_\parallel(z, 0) + i\omega\tilde{\phi}^+_\parallel(z, \omega) +  a_\parallel(z)}{2 \varepsilon_0 k_B T} \, , \label{eq:eps_parallel} \\
		\frac{\chi_\perp(z, \omega)}{\chi_\perp(z, \omega) + 1} &= \frac{\phi_\perp(z, 0) + i \omega \tilde{\phi}^+_\perp(z,\omega) +  a_\perp(z)}{\varepsilon_0 k_B T +( \Phi_\perp(0) + i \omega \tilde{\Phi}^+_\perp(\omega) +  A_\perp ) / V_\mathrm{sys}}\,, \label{eq:eps_perp}
\end{align}
where $\tilde{f}(\omega) = \int\limits_{-\infty}^\infty \mathrm{d}t\, e^{i \omega t}f(t)$ denotes the one-sided Fourier transform, and we define the the polarization correlation functions
\begin{align}
		\phi_\alpha(z, t) =& \left\langle \mathbf{m}_\alpha(z, 0) \cdot \mathbf{M}_\alpha(t)\right\rangle - \left\langle \mathbf{m}_\alpha(z)\right\rangle \cdot \left\langle \mathbf{M}_\alpha\right\rangle\,,  \\ 
		\Phi_\alpha(t) =& \left\langle \mathbf{M}_\alpha(0) \cdot \mathbf{M}_\alpha(t)\right\rangle - \left\langle \mathbf{M}_\alpha\right\rangle \cdot \left\langle \mathbf{M}_\alpha\right\rangle\, ,
		\label{eq:pol_corr}
	\end{align}
for $\alpha=\parallel,\perp$ with $\mathbf{m}_\alpha(z,t)$ being the polarization density profile parallel ($\alpha=\,\parallel$) or perpendicular ($\alpha=\,\perp$) to the interface, and $\mathbf{M}_\alpha(t) = \int\limits_{V_\mathrm{sys}} \mathrm{d}\mathbf{r}\, \mathbf{m}_\alpha(z, t)$ the total dipole moment of the simulation box with a volume of $V_\mathrm{sys}$. Since within the Born-Oppenheimer approximation, DFT-MD simulations exhibit instantaneous polarizability, the position-dependent electronic polarizability $a_\alpha(z)$ and the total system electronic polarizability $ A_\alpha = \int\limits_{V_\mathrm{sys}} \mathrm{d}\mathbf{r}\,  a_\alpha(z)$ in Eqs.~\ref{eq:eps_parallel}\,-\,\ref{eq:eps_perp} have to be taken into account.

To relate our simulations to spectroscopic experiments we could obtain the parallel and perpendicular components of the position-dependent dielectric spectrum $\chi_\parallel(z,\omega)$ and $\chi_\perp(z, \omega)$ using Eqs.~\ref{eq:eps_parallel} and \ref{eq:eps_perp}. However, given the limited amount of sampling one can obtain from DFT-MD simulations, the spatially resolved spectra have a too low signal-to noise ratio to provide insightful information. Instead, we determine the spectrum of the water in the first hydration layer of the interface according to 
\begin{align}
\begin{split}
\chi_\parallel^\mathrm{fhl}(\omega) = \frac{A_\mathrm{sys}}{n_\mathrm{w}^\mathrm{fhl}v_\mathrm{w}} \times \\ \int\limits_{-\infty}^{z_\mathrm{fhl}} \mathrm{d}z\, \frac{\phi^\mathrm{w}_\parallel(z, 0) + i\omega\tilde{\phi}^\mathrm{w,+}_\parallel(z,\omega) + a^\mathrm{w}_\parallel(z)}{2\varepsilon_0k_\mathrm{B}T}\,, \label{eq:sams:chi_fhl_par}
\end{split}\\
\begin{split}
\frac{\chi_\perp^\mathrm{fhl}(\omega)}{\chi_\perp^\mathrm{fhl}(\omega)+1} = \frac{A_\mathrm{sys}}{n_\mathrm{w}^\mathrm{fhl}v_\mathrm{w}} \times \\ \int\limits_{-\infty}^{z_\mathrm{fhl}} \mathrm{d}z\, \frac{\phi^\mathrm{w}_\perp(z, 0) + i\omega\tilde{\phi}^\mathrm{w,+}_\perp(z,\omega) + a^\mathrm{w}_\perp(z)}{\varepsilon_0k_\mathrm{B}T + (\Phi(0) + i\omega\tilde{\Phi}_\perp^+(\omega)) + A_\perp)/V_\mathrm{box}}\,.\end{split} \label{eq:sams:chi_fhl_perp}
\end{align}
For this, we decompose the system's polarization in terms of a water and a surface contribution $\tilde{m}_\alpha(z, \omega) = m_\alpha^\mathrm{w}(z, t) + m_\alpha^\mathrm{surf}(z,t)$ with $\alpha=\parallel,\perp$, leading to the correlation functions
\begin{align}
    \phi_\alpha^\mathrm{w}(z,t) = \left\langle m^\mathrm{w}_\alpha(z,t)M_\alpha(t) \right\rangle\, .
\end{align}
Accordingly, $a_\parallel^\mathrm{w}(z)$ and  $a_\perp^\mathrm{w}(z)$ denote the position-dependent polarizabilities of the water molecules. The applicability of this approach has been shown previously \cite{becker2024interfacial}. In practice, we use maximally localized Wannier functions to extract the dipole moment of each water molecule $\boldsymbol{\mu}_i^\mathrm{w}(t)$ and approximate the water polarization density by binning with respect to the center of mass position of the water molecules $z^\mathrm{COM}$, using
\begin{align}
    m^\mathrm{w}_\alpha(z,t) \approx \sum\limits \delta(z - z^\mathrm{COM}_i(t)) \mu_{i\alpha}^\mathrm{w}(t)\, .
\end{align}
Finally, we normalize the spectral contributions of the hydration water to the average number of molecules found in the first hydration layer $n_\mathrm{w}^\mathrm{fhl}$, which defines the effective length of the first hydration layer 
\begin{align}
    L_\mathrm{w}^\mathrm{fhl} = \frac{n_\mathrm{w}^\mathrm{fhl} v_\mathrm{w}}{A_\mathrm{sys}}\, .
\end{align}
Here, $A_\mathrm{sys}$ denotes the lateral area of the simulation cell, and $v_\mathrm{w}=0.03036~\mathrm{nm}^3$ is the specific volume of water molecules in bulk  water. 

\begin{figure}[ht]
    \centering
    \includegraphics[width=1.0\linewidth]{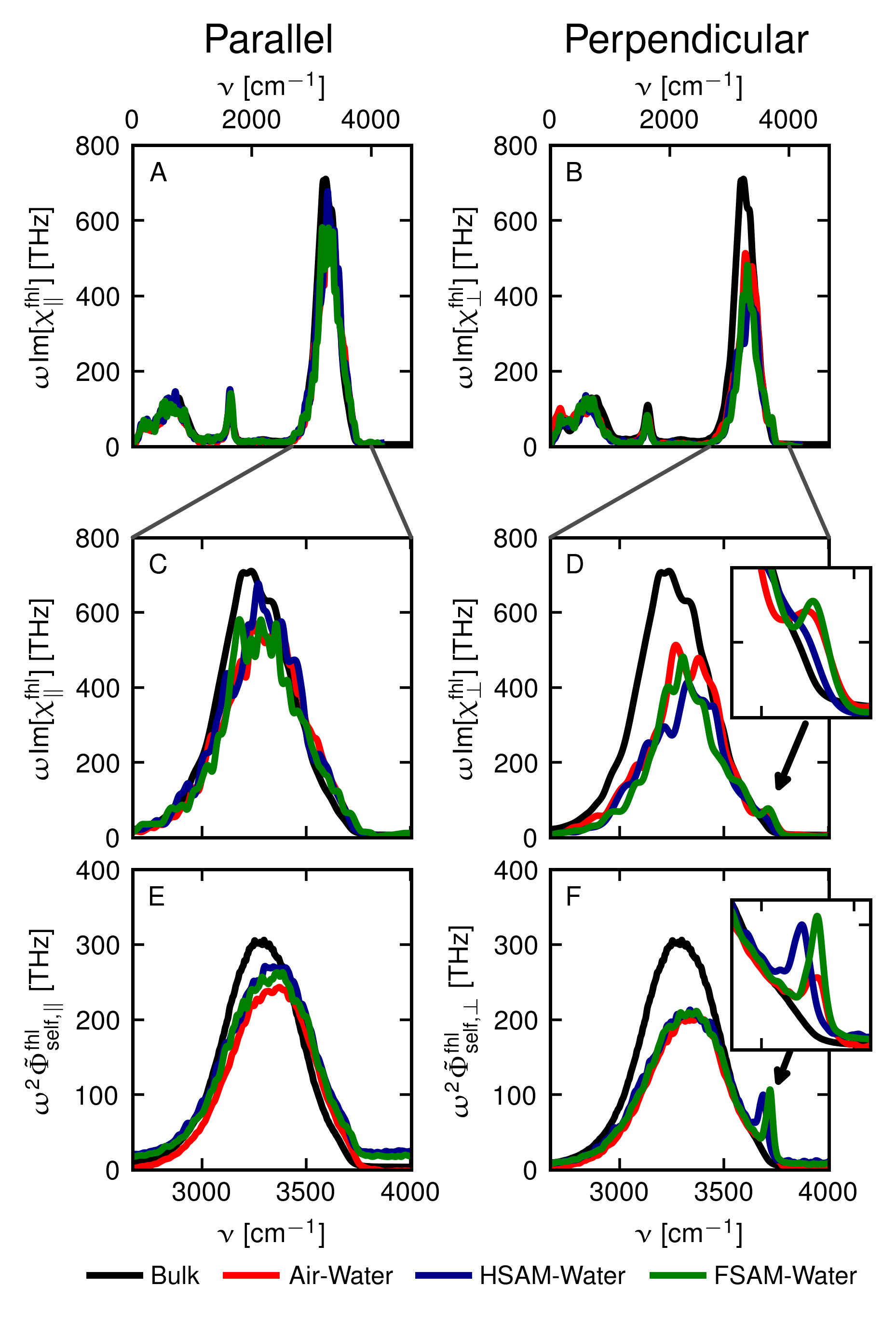}
    \caption{Vibrational spectra of water in the first hydration shell at various interfaces. \textbf{A}-\textbf{B}: Parallel and perpendicular component of interfacial spectra according to Eqs.~\ref{eq:sams:chi_fhl_par}-\ref{eq:sams:chi_fhl_perp}. \textbf{C}-\textbf{D}: Close up of the OH stretch vibrational region. \textbf{E}-\textbf{F}: Self contributions to interfacial spectra according to Eq.~\ref{eq:self_corr} (methods). Arrows indicate the signal corresponding to the vibration of free OH groups.}
    \label{fig:sams:ir_spectra}
\end{figure}

Parallel and perpendicular components of the absorption spectrum $\omega\mathrm{Im}[\chi^\mathrm{fhl}]$ of interfacial water are shown in Fig.~\ref{fig:sams:ir_spectra}\,A-B. They highlight the anisotropic nature of interfacial water: Over the entire frequency range, the parallel and perpendicular components differ significantly from each other as well as from the bulk spectrum. The four major peaks of the water spectrum in the THz regime, the hydrogen bond stretch peak at 150~cm$^{-1}$, the libration peak at 1000~cm$^{-1}$, the OH bend peak at 1700~cm$^{-1}$ and the OH stretch peak at 3300~cm$^{-1}$, give complementary information on the hydrogen bond network at hydrophobic interfaces \cite{LeCaer2011, seki2020bending, pezzotti2021molecular, lehmann2025beyond}. The anisotropy of the linear interfacial absorption spectrum is most striking in the OH stretch region (Fig.~\ref{fig:sams:ir_spectra}\,C-D), which we focus on in this work. It is experimentally the best characterized and, besides being indicative of hydrogen bonding, contains contributions from free OH groups that do not participate in hydrogen bonding but interact directly with the surface, thereby encoding critical information about water–surface interactions.

The position and line shape of the OH stretch peak of water provide valuable insights into the hydrogen bond network of water, as both parameters are known to depend on hydrogen bond strength \cite{nakamoto1955stretching}. It has been argued that in liquid water, the OH stretch peak can be decomposed into contributions from populations with different numbers of hydrogen bonds \cite{bergonzi2014gibbs}, allowing for the tracking of the hydrogen bonding structure. While the uniqueness of such a decomposition based on fitting experimental spectra is questionable due to the large spectral overlap of different contributions, but also the intricate nature of hydrogen bonding environments \cite{auer2007hydrogen, tainter2013hydrogen}, the overall interpretation that a higher frequency of the OH stretch vibration is related to a lower degree of hydrogen-bonding and a lower frequency to a higher degree of hydrogen-bonding is largely accepted \cite{calegari2023probing, wang2025quantifying}. Here, we determine the center frequency $\omega_\mathrm{g}$ of the OH stretch band by fitting to a single Gaussian function
\begin{align}
    \omega\chi''(\omega) \propto e^{-\frac{(\omega-\omega_\mathrm{g})}{2\sigma_g^2}}\, .
    \label{eq:sams:gaussian}
\end{align}
The resulting fits are shown in Sec.~S4 of the supplementary material and the corresponding fit parameters are listed in Tab.~\ref{tab:sams:fit_params}. They exhibit an overall blue shift of the OH stretch peak at all simulated interfaces, the air-water the FSAM-water and HSAM-water interfaces, with respect to the bulk of approximately 30-60~cm$^{-1}$ for both the parallel and perpendicular components. According to the interpretation we just laid out, this indicates, consistent with the hydrogen-bond profiles in Fig.~\ref{fig:orientations}\,K, a lower degree of hydrogen-bonding in the first layer of water at all interfaces. This finding is in agreement with experimental interpretations of IR and Raman spectroscopy in various confinement geometries \cite{LeCaer2011, Knight2019, calegari2023probing, wang2025quantifying, martins2026sub}.

\begin{table}[ht]
    \centering
    \caption{Fit parameters for the OH stretch feature of the different spectra shown in Fig.~\ref{fig:sams:ir_spectra} and Fig.~\ref{fig:fig_4}.  All fits are shown in Sec.~S4 of the supplementary material.}
    \begin{tabular}{clccccc}
        \toprule 
         &&& $\frac{\omega_\mathrm{g}}{2\pi}$ [cm$^{-1}$] & $\sigma$ [cm$^{-1}$] & $\frac{\omega_\mathrm{f}}{2\pi}$ [cm$^{-1}$] & $\Delta$ [cm$^{-1}$] \\ \midrule
         $\omega\chi''$&Bulk &  & 3256 & 175 & - & - \\
         $\omega\chi''_\mathrm{fhl}$ &Air-Water & $\parallel$ & 3303 & 195 & - & - \\ 
         && $\perp$ &  3323 & 173 & - & - \\ 
         &HSAM-Water & $\parallel$ & 3293 & 192 & - & - \\ 
         && $\perp$ & 3321 & 188 & - & - \\
         &FSAM-Water & $\parallel$ & 3286 & 190 & - & - \\ 
         && $\perp$ & 3312 & 165 & - & - \\
         \midrule 
        $\omega^2\tilde\Phi^\mathrm{self}$ & Bulk & & 3289 & 182 & - & - \\
        $\omega^2\tilde\Phi^\mathrm{self}_\mathrm{fhl}$ & Air-Water & $\parallel$ & 3337 & 195  & - & - \\
        && $\perp$ & 3326 & 187 & 3717 & 35 \\
        &HSAM-Water & $\parallel$ & 3328 & 218 & - & - \\
        & & $\perp$ & 3320 & 202 & 3687 & 29 \\
        &FSAM-Water & $\parallel$ & 3328 & 214& -& - \\
        & & $\perp$ & 3321 & 194 & 3720 & 27 \\ \midrule $\omega^2\tilde{S}_\mathrm{OH}^\mathrm{free}$ & Air-Water & &  -& - & 3718 & 33 \\ 
         & HSAM-Water & & -& - & 3689 & 34   \\
         & FSAM-Water & & - & - & 3720 & 27 \\\bottomrule
    \end{tabular}
    \label{tab:sams:fit_params}
\end{table}

Computational IR absorption spectra are often challenging to interpret. In hydrogen-bonded systems, the OH stretching vibrations are strongly correlated, leading to significant collective features in the spectrum that obscure the spectroscopic fingerprints of specific molecular motions \cite{carlson2020exploring}. Moreover, these collective contributions are typically noisy and converge slowly in DFT-MD simulations, which provide only limited trajectory lengths. To interpret the OH stretch spectrum on a molecular level, we determine the single-molecular contribution $\omega^2\tilde{\Phi}_\mathrm{self}^\mathrm{fhl}$ to the absorption spectrum in the first hydration layer using the decomposition in Eq.~\ref{eq:self_corr} (methods). Results are shown in Fig.~\ref{fig:sams:ir_spectra}\,E-F. The parallel component of the single-molecular contributions is qualitative similar to the bulk spectrum. In the perpendicular component however, we observe the emergence of a narrow peak at a frequency higher than the usual OH stretch band, which we identify as the spectral signature of free (non hydrogen-bonded) OH groups.  Given the limited sampling of our simulations, these features exhibit intensities too small to be discernible against the noise of the full spectra, however as shown later, they are experimentally accessible.

In comparison to hydrogen-bonded water molecules, the chemical environments of free OH bonds are rather uniform, giving rise to the narrow line width of the corresponding vibrational peak. The spectral properties of free OH groups can be modeled in terms of a single damped harmonic oscillator subject to a driving random force
\begin{align}
    \ddot{x}(t) + \frac{\gamma}{m} \dot{x}(t) + \omega_\mathrm{f}^2x(t) = \frac{1}{m}F_\mathrm{R}(t)\,, \label{eq:sams:damped_harmonic}
\end{align}
where $m$ is the mass, $\gamma$ a friction coefficient, $\omega_\mathrm{f}$ the resonance frequency of the undamped oscillator and $F_\mathrm{R}$ the random force. The power spectral density of such an oscillator $S_x(\omega)$, also called vibrational density of states, can be equated to \cite{}
\begin{align}
    \gamma\omega^2\tilde{S}_{x}(\omega)= \frac{\Delta^2 \omega^2}{(\omega_\mathrm{f}^2-\omega^2)^2 + \Delta^2\omega^2}\, . \label{eq:sams:lorentzian}
\end{align}

This function is of Lorentzian shape, characterized by the cen  ter frequency $\omega_\mathrm{f}$ and the full width at half maximum $\Delta=\frac{\gamma}{m}$. By fitting the self-spectra of the bulk and the parallel component in Fig.~\ref{fig:sams:ir_spectra}\,E with a single Gaussian (Eq.~\ref{eq:sams:gaussian}), and the perpendicular component in Fig.~\ref{fig:sams:ir_spectra}\,F with one Gaussian and one Lorentzian function (Eq.~\ref{eq:sams:lorentzian}), we recover blue shifts of the Gaussian peak of interfacial water relative to bulk that are consistent with those found in the full absorption spectrum $\omega\chi''(\omega)$. All fits are shown in Sec.~S4 of the supplementary material, and the corresponding parameters are reported in Tab.~\ref{tab:sams:fit_params}. This result indicates that the blue shifts observed in Fig.~\ref{fig:sams:ir_spectra}\,A–B originate from modifications of the vibrational motion of individual OH bonds, rather than from collective effects. Additionally, the fits show that a single Lorentzian function is a good approximation for the line shape of the spectral contribution of free OH groups.

When comparing the fitted parameters in Tab.~\ref{tab:sams:fit_params} for the OH stretch spectra of water across the different interfaces, the following picture emerges: at all three simulated interfaces, the position of the hydrogen-bonded main OH peak $\omega_\mathrm{g}$ is blue shifted with respect to bulk by 30 - 60~cm$^{-1}$, both in the parallel and perpendicular components. For all considered spectral observables ($\chi_\parallel$, $\chi_\perp$ , $\tilde{\Phi}^\mathrm{self}_\parallel$ and $\tilde{\Phi}^\mathrm{self}_\perp$) blue shifts are the largest at the air-water interface. At the HSAM and FSAM interfaces, the blue shifts are somewhat weaker than at the air–water interface. Notably, the different spectral observables do not fully agree: the overall spectrum $\chi(\omega)$ exhibits stronger blue shifts at the HSAM interface than at the FSAM interface, whereas the self-spectra $\tilde{\Phi}^\mathrm{^self}(\omega)$ show identical shifts at both. Interestingly, the frequency shifts of $\omega_\mathrm{g}$ correlate with the maximum peak heights in the density profiles shown in Fig.~\ref{fig:orientations}\,A-C. This indicates tighter hydrogen bonding in the first hydration layer of HSAMs and FSAMs compared to the air-water interface.

\subsection{Free OH groups as spectroscopic fingerprints of water-surface interactions}

The frequencies of the free OH peak as fitted from $\tilde{\Phi}^\mathrm{self}(\omega)$ (See Tab.~\ref{tab:sams:fit_params}) follow a different ordering $\omega_\mathrm{f}^\mathrm{FSAM}\gtrsim \omega_\mathrm{f}^\mathrm{air}>\omega_\mathrm{f}^\mathrm{HSAM}$. Notably, the free OH peak is found at the same, or even slightly higher, frequency for the FSAM-water interface compared to the air-water interface. This challenges the conventional interpretation that this frequency can serve as a microscopic indicator of hydrophobicity \cite{pezzotti2021molecular}. According to that interpretation, the free OH frequency is dominated by the vibrational Stark effect from electrostatic interactions with the surface: more hydrophilic surfaces cause stronger red-shifts relative to the gas-phase OH stretch. At the air–water interface, lacking any surface interaction, one would expect the maximum free OH frequency. Our results show this relation is not universal, indicating that dispersive interactions or water–water interactions also influence the free OH vibrational spectra.

\begin{figure}
    \centering
    \includegraphics[width=1.0\linewidth]{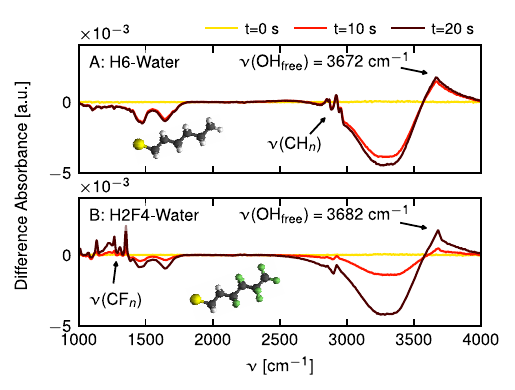}
    \caption{Difference SEIRA spectra of the water-stretching region after adsorption of a monolayer of (A) 1-hexanethiol and (B) perfluorohexanethiol. Spectral features from the gold–water interface and the liquid bulk water displaced by the monolayer volume appear as negative bands. Vibrational features from the monolayer and the newly created interfacial water structure appear as positive bands.}
    \label{fig:experimental}
\end{figure}

To experimentally test our predictions regarding the frequency of free OH vibrations, we conduct surface enhanced infrared absorption spectroscopy experiments at the HSAM-water and FSAM-water interfaces. As model molecules, we used 1-hexanethiol and 1,1,2,2-tetrahydrogen-perfluorohexane-1-thiol, both with a molecular length of six carbon atoms (see Fig.~\ref{fig:experimental} insets). The interfacial water structure was measured by depositing a monolayer of the HSAM or FSAM compound directly from water. To increase the solubility of these compounds, we used an alkaline \ce{Na2CO3} buffer (pH = 9.5) and kept the concentration low during deposition (1.5 mM). First, a background spectrum was recorded by covering the bare gold surface with the alkaline buffer. The surface enhancement factor characteristic of SEIRAS is restricted to a distance of about $\approx$100 nm from the gold surface \cite{tseng2023plasmon}. This allows us to observe the vibrational response of molecules within this distance and removes the strong contribution from the bulk water above the monolayer. The electric field in SEIRAS is polarized perpendicular to the gold surface, enhancing vibrational modes with transition dipole moments oriented in this direction. This makes SEIRAS particularly suitable for detecting signatures of free OH groups \cite{ataka1996potential}, which were observed only in the perpendicular IR spectra (Fig.~\ref{fig:sams:ir_spectra}\,D,F). After adding the thiol solution, the gold substrate gets quickly covered with the HSAM/FSAM molecules. As the SAM forms, water molecules are displaced away from the gold interface. This removes the spectroscopic signature from the gold-water interface and the contribution from the liquid water volume occupied by the monolayer. After 20 s, surface coverage reaches saturation, which is observed by a stabilization in the intensity of the $\nu(\mathrm{CH_{n}})$ bands at $\sim$2950 cm$^{-1}$ and the $\nu(\mathrm{CF_{n}})$ bands at $\sim$1250 cm$^{-1}$ of the HSAM and FSAM \cite{cruz2025infrared}, respectively (Fig.~\ref{fig:experimental}). In both experiments, the broad O-H stretching vibrational band of water is observed at 3285 cm$^{-1}$. The fact that the band appears negative relates to the displacement of liquid bulk water by the monolayer volume and from the lost water structure at the gold interface. At higher frequencies, a sharp positive band is observed at 3682 cm$^{-1}$ for the FSAM and at 3672 cm$^{-1}$ for the HSAM, characteristic frequencies of free O-H groups of the newly created water structure at the water-monolayer interface. This dangling-water band appears notably sharper in the FSAM. These results confirm our simulation predictions: free OH vibrations are blue shifted at the FSAM-water interface with respect to the HSAM-water interface. 



\begin{figure}[ht]
    \centering
    \includegraphics[width=1.0\linewidth]{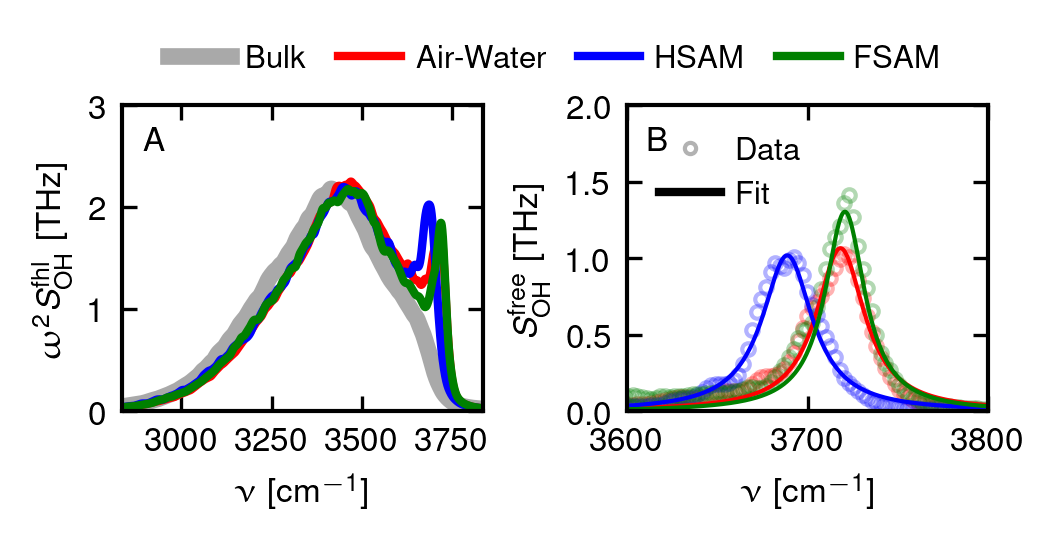}
    \caption{\textbf{A}: Power spectrum of the OH bond length in the first hydration layer of different interfaces compared to bulk water. \textbf{B}: Power spectrum of the OH bond length for free OH groups using the criterion ($z_\mathrm{COM}<z_\mathrm{WCI}+0.9~\mathrm{\AA}$, $\theta_\mathrm{OH}>120^\circ$). Fits to a single Lorentzian function using Eq.~\ref{eq:sams:lorentzian} are indicated as solid lines.}
    \label{fig:fig_4}
\end{figure}

To analyze how the differences in the spectral properties of free OH groups relate to the molecular structure and dynamics of water at various interfaces, we establish a geometric criterion to identify free OH groups. Following Tang et al. \cite{tang2018definition}, we maximize the spectral overlap between the high-frequency peak of the interfacial water vibrational spectra and the power spectral density of the putative free OH population. Unlike their distance-based water–water criterion \cite{tang2018definition}, ours relies on the geometrical parameters in Fig.~\ref{fig:orientations}: an OH group is deemed free if it belongs to a water molecule whose center-of-mass position fulfills $z_\mathrm{COM}<z_\mathrm{free}=z_\mathrm{WCI}+0.9~\mathrm{\AA}$ and the corresponding OH group forms an angle of more than $\theta_\mathrm{OH}>120^\circ$ with the interfacial normal. We discuss this definition of free OH groups in detail in Sec.~S5 of the supplementary material. The key quantities, namely the power spectral density of OH groups of interfacial water molecules used to construct the geometric criterion and the power spectral density of the resulting population of free OH groups, are shown in Fig.~\ref{fig:fig_4}, panels A and B, respectively. The resulting power spectral density of free OH groups, which shows contributions only in the high frequency regime is centered around approximately 3700~cm$^{-1}$, and can be accurately fitted with the Lorentzian line shape function in Eq.~\ref{eq:sams:lorentzian}. The resulting fit parameters, which are given  in Tab.~\ref{tab:sams:fit_params}, reproduce the frequencies of free OH groups found from fitting $\omega^2\tilde\Phi_\mathrm{fhl}^\mathrm{self}$ in the high frequency region across all interfaces.

\begin{figure*}[ht]
    \centering
    \includegraphics[width=0.85\linewidth]{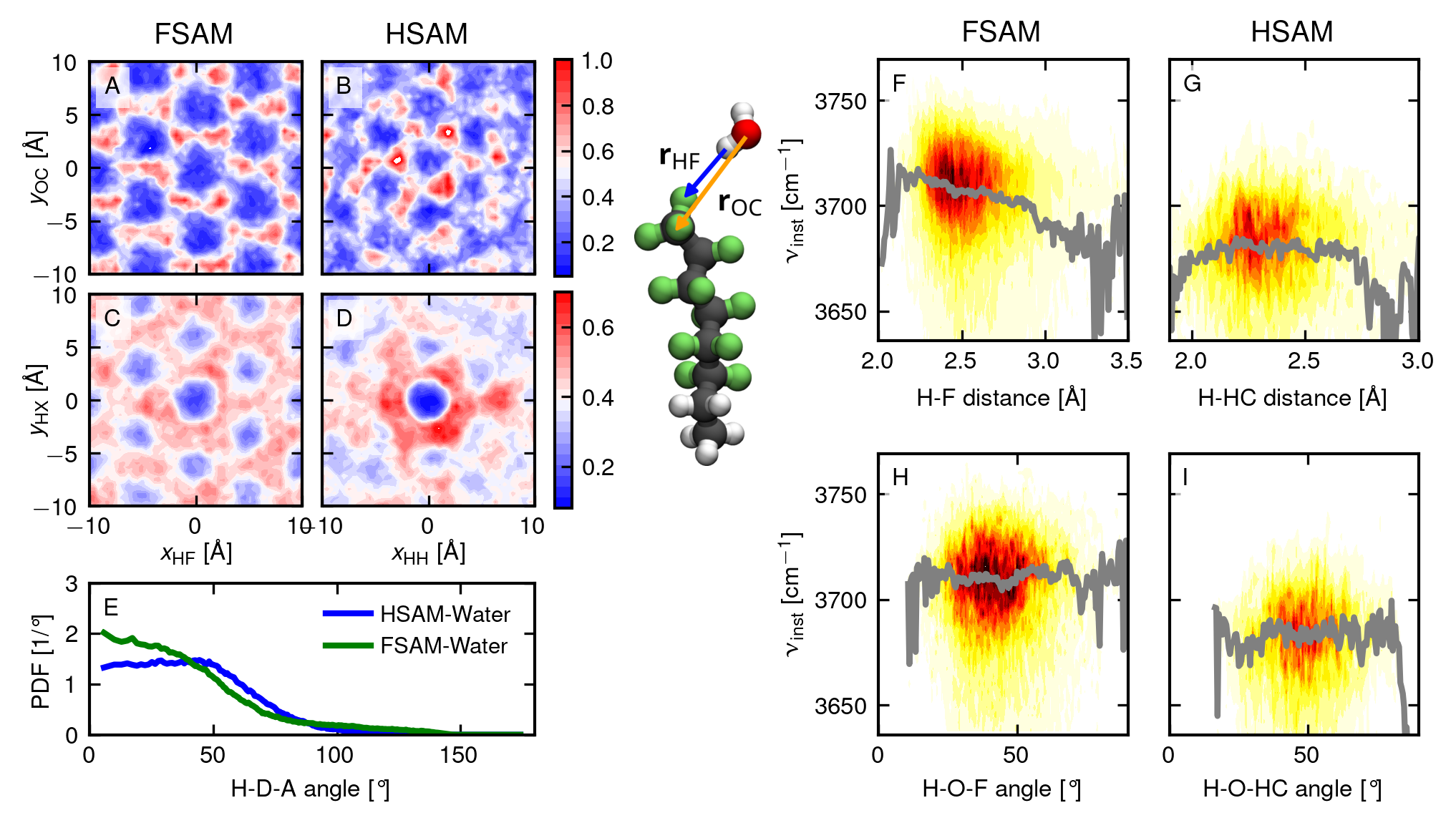}
    \caption{\textbf{A-D}: Spatial distribution of free OH groups on a SAM. We show the probability distribution as a function of the x and y components of the $\mathbf{r}_\mathrm{HX}$, $\mathrm{X}\in(\mathrm{F,H})$ and the $\mathbf{r}_\mathrm{OC}$ vectors indicated on the right side of the heat maps. Probability distributions are averaged over all HX and OC pairs. \textbf{E}: Probability distribution of the angle between a free OH group's bond direction (H - donor hydrogen of free OH group, D - donor oxygen of free OH group) and the vector connecting its oxygen atom to the nearest hydrogen or fluorine atom on the SAM surface (A - acceptor atom). \textbf{F}-\textbf{I}: Probability density plot from correlating geometrical descriptors of free OH group surface interactions with the instantaneous frequency of free OH vibrations found by performing time-frequency analysis according to Eq.~\ref{eq:instantaneous_freq} (see methods). Geometric descriptors (H-X distances and H-O-X angles) are averaged over the same windows used in time-frequency analysis to determine instantaneous frequencies (see methods). For better visibility the probability density is averaged according to Eq.~\ref{eq:local_average} resulting in the gray lines.}
    \label{fig:sam:free_OH}
\end{figure*}

Using our criterion, we conclude that the number of free OH groups at the three different interfaces is rather similar. We find an average of 1.17, 1.20, and 1.25 free OH groups per $\mathrm{nm}^2$ of interfacial area at the air-water, HSAM-water, and FSAM-water interfaces, respectively. The similar populations of free OH groups near fluorinated (FSAM) and non-fluorinated (HSAM) SAMs contrasts with reports for the hydration shells of small fluorinated organic molecules, where free OH bonds are significantly more prevalent than in their non-fluorinated counterparts \cite{robalo2019hydrophobic}. This illustrates the different hydration properties of surfaces and small molecules: The polarity of C-F bonds majorly influences the hydration shell of small molecules. At fluorinated surfaces however, the individual C-F bonds are not accessible to the water, and the polarity of C-F bonds is not felt. Instead, dispersive interactions dominate the interactions of water with fluorinated SAMs, consistent with conclusions drawn from  previous FF-MD simulations \cite{carlson2021hydrophobicity}.

To illustrate this further, we map out the spatial distribution of free OH groups in Fig.~\ref{fig:sam:free_OH}\,A-D. In particular, for the FSAM-water interface, we observe a clear spatial ordering that mirrors the hexagonal pattern of SAM molecules. The data indicate that free OH groups are not positioned directly on top of SAM molecules, which would signify strong interactions between fluorine and hydrogen atoms; rather, they reside on top of the void spaces between them. It is noteworthy that water molecules do not actually penetrate the void spaces between the SAM molecules, as evidenced by the substantial depletion layer discussed in Fig.~\ref{eq:sams:depletion_layer}. Also, the angular distribution of free OH groups in Fig.~\ref{fig:sam:free_OH}\,E demonstrates that OH groups exhibit only a slightly stronger affinity for orienting towards fluorine atoms as compared to hydrogen atoms. 

These geometrical considerations can be linked to the vibrational properties of free OH groups. We therefore compute the instantaneous vibrational frequency $\nu_\mathrm{inst}(t)$ of free OH groups via Gabor transform of their bond lengths (see Methods for details). To assess the impact of the local environment on their spectral properties, we correlate the instantaneous vibrational frequency $\nu_\mathrm{inst}(t)$ with the geometrical descriptors of Fig.~\ref{fig:sam:free_OH}\,A-E, the distance between free OH hydrogen atoms and the closest fluorine atom or hydrogen atom of the SAM (Fig.~\ref{fig:sam:free_OH}\,F-G) as well as the angle at which the free OH group points towards the closest SAM atom (H-O-F and H-O-HC angle in Fig.~\ref{fig:sam:free_OH}\,H-I). The only geometrical descriptor which shows significant correlation to the instantaneous free OH vibrational spectroscopy is the distance between free OH hydrogen atoms and FSAM fluorine atoms (Fig.~\ref{fig:sam:free_OH}\,F). Strikingly, the relation we observe stands in contrast to the expected behavior from the vibrational Stark effect: the closer the H-F distance, the higher the observed vibrational frequency. The same is not observed for the HSAM-water interface for the distance between water hydrogen atoms and SAM hydrogen atoms H-HC distance (Fig.~\ref{fig:sam:free_OH}\,G). Interestingly, also the angle that the free OH groups form to the closest fluorine and hydrogen atoms exhibits no correlation to the instantaneous frequency of free OH groups, as would be expected if H$\cdots$F hydrogen bonds were formed between free OH groups and FSAMs.

\subsection{The line shape of free OH spectra reveal a slowdown of water interfacial dynamics at fluorinated SAMs}

\begin{figure}[ht]
    \centering
    \includegraphics[width=1.0\linewidth]{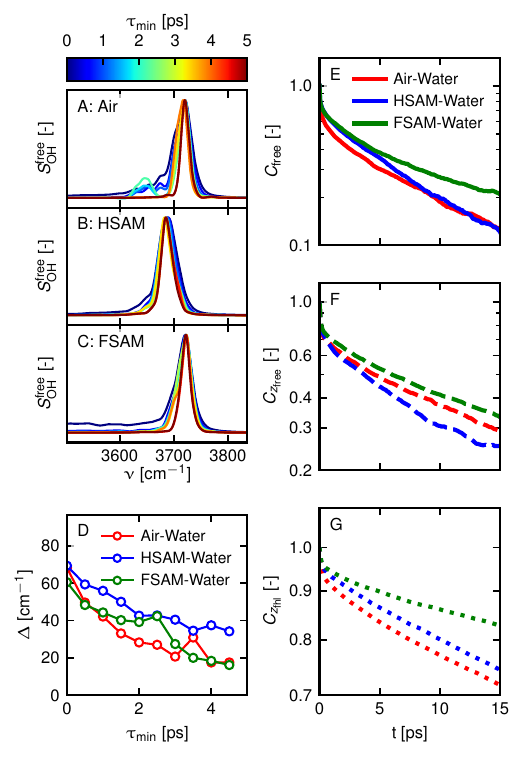}
    \caption{\textbf{A}-\textbf{C}: Power spectra of free OH bond lengths, for different minimum  residence times $\tau_\mathrm{min}$. \textbf{D}: Line widths of the power spectra in panels A-C, found by fitting to single Lorentzian functions. \textbf{E}-\textbf{G}: Autocorrelation functions of the indicator functions in Eqs.~\ref{eq:sams:indicator1}-\ref{eq:sams:indicator2} indicating lifetimes of free OH groups, and the residence times of water molecules in the regions $z_\mathrm{COM}<z_\mathrm{free}$ and $z_\mathrm{COM}<z_\mathrm{fhl}$. }
    \label{fig:sams:lifetimes}
\end{figure}

Finally, we note that not only the positions of the free OH vibrational peaks but also their lineshapes encode information about water–surface interactions. When fitted directly from the $\omega^2\Phi_\mathrm{fhl}^\mathrm{self}$, we find that the line width of the free OH peak is  the largest at the air-water interface ($\Delta$ = 35~cm$^{-1}$ in Tab.~\ref{tab:sams:fit_params}) compared to that at the HSAM-water ($\Delta$ = 29~cm$^{-1}$) and FSAM-water ($\Delta$ = 27~cm$^{-1}$) interfaces. Analysis of the power spectra of free OH bond lengths using the criterion from Fig.~\ref{fig:fig_4} yields similar line widths.

The line widths of a Lorentzian peak can be interpreted as the lifetime of the underlying vibration. However, the vibration of free OH groups cannot be described by a single lifetime, as there are at least two different processes involved: (i) the vibration of a free OH group can de-phase due to interactions with the surface and internal coupling within the same molecule, giving rise to an intrinsic lifetime. (ii) the water molecule can reorient or translate, such that the free OH is converted to a hydrogen-bonded OH shifting its frequency. To isolate the intrinsic vibrational lifetime from residence effects, we compute power spectra of OH bond lengths for groups that remain classified as 'free' continuously for times exceeding $\tau_\mathrm{min}$. This filters out free OH groups that move too rapidly (losing their free status), revealing whether intrinsic dephasing or reorientation dominates the observed line shape. The resulting power spectra for different $\tau_\mathrm{min}$ are shown in Fig.~\ref{fig:sams:lifetimes}\,A-C. They reveal that taking into account only long-lived OH groups (large $\tau_\mathrm{min}$) results in a spectrum with a significantly narrower line width. This suggests that the line widths of free OH spectral features are determined by the residence lifetime of free OH groups, rather than their intrinsic vibrational lifetimes. We fit the corresponding free OH spectra for different $\tau_\mathrm{min}$ by a single Lorentzian function each and determine the corresponding line widths, see Fig.~\ref{fig:sams:lifetimes}\,D. The line widths of OH vibrations at the air-water and FSAM-water interfaces decrease from $\Delta\approx60$~cm$^{-1}$ when taking all free OH groups into account ($\tau_\mathrm{min}=0$~ps) to a value of $\Delta\approx15$~cm$^{-1}$ for long-lived free OH groups ($\tau_\mathrm{min}=5$~ps). The saturation of the linewidth with respect to $\tau_\mathrm{min}$ marks the intrinsic dephasing time, which for free OH groups exceeds the residence lifetime by a factor of around 4.  Interestingly, the plateau value remains higher ($\Delta\approx 30$~cm$^{-1}$) at the HSAM–water interface.

The differences in line widths of the free OH peaks among the air–water, HSAM–water, and FSAM–water interfaces, observed in Fig.~\ref{fig:sams:ir_spectra}\,F and quantified in Tab.~\ref{tab:sams:fit_params}, reflect varying rotational dynamics of water molecules. We quantify these dynamics in terms of the correlation function $C_\mathrm{free}(t)$, $C_{z_\mathrm{free}}(t)$ and $C_{z_\mathrm{fhl}}(t)$ of the indicator functions
\begin{align}
    \theta_\mathrm{free} =\begin{cases}
        1 \qquad \mathrm{if\;free\;OH}\,,\\
        0 \qquad \mathrm{else}\,
    \end{cases} \label{eq:sams:indicator1}
\end{align}
and
\begin{align}
    \theta_{z_\mathrm{free/fhl}} =\begin{cases}
        1 \qquad \mathrm{if}\;z_\mathrm{COM} < z_\mathrm{free/fhl} \,\\
        0 \qquad \mathrm{else}\,.
    \end{cases} \label{eq:sams:indicator2}
\end{align}
Results are shown in Fig.~\ref{fig:sams:lifetimes}\,E-G. Indeed, the lifetime of free OH groups is longer at the FSAM-water interface than at the air-water and HSAM-water interfaces, as indicated by the slower decay of the $C_\mathrm{free}(t)$ correlation function in Fig.~\ref{fig:sams:lifetimes}\,E. Since our criterion for free OH groups involves both $z_\mathrm{COM}<z_\mathrm{free}$ and $\theta_\mathrm{OH}<120~^\circ$, comparing $C_\mathrm{free}(t)$ (Fig.~\ref{fig:sams:lifetimes}\,E) with $C_{z_\mathrm{free}}(t)$ (Fig.~\ref{fig:sams:lifetimes}\,F) reveals that free OH lifetimes are determined approximately equally by translational motion reorientation. Moreover, not only are free OH dynamics slowed at the FSAM–water interface relative to the other two, but the translational motion of hydration water in general is also decelerated, as shown by $C_{z_\mathrm{fhl}}$ in Fig.~\ref{fig:sams:lifetimes}\,G. 

Interfacial water at strongly hydrophobic surfaces is known to exhibit slower reorientational dynamics than at surfaces of intermediate hydrophobicity, primarily due to enhanced water–water interactions, similar to the slowdown observed around hydrophobic solutes \cite{rezus2007observation, romero2009effect, stirnemann2011non}. Since the air–water interface represents the most hydrophobic limit, the pronounced slowdown observed at HSAM and FSAM surfaces must involve an additional mechanism, most likely a retardation arising from surface–water interactions. Such a process is typically associated with strongly polar, and thus hydrophilic, interfaces, whose interpretation is also more consistent with the magnitude of the observed slowdown  \cite{stirnemann2011non, gekle2012anisotropy}.  This leads to the intriguing conclusion that fluorinated SAMs, despite being macroscopically more hydrophobic than their unfluorinated counterparts, exhibit also some spectroscopic characteristics usually associated with increased hydrophilicity.

\section{Conclusion}

We present large-scale DFT-MD simulations of water at self-assembled monolayers of fluorinated and non-fluorinated alkanes, comparing these to the prototypical hydrophobic interface, the air-water interface. Although fluorinated and non-fluorinated alkanes are chemically distinct, particularly considering that C-F bonds are more polar than C-H bonds, we observe strikingly similar water organization at all three interfaces in terms of density layering and molecular orientation. This finding corroborates previous theoretical studies predicting that the dominant interaction between fluorinated surfaces and water is dispersive \cite{carlson2024modeling}, which explains the strong hydrophobicity of fluorinated surfaces. 

To make our observations of the microscopic water structure experimentally testable, we discuss the characteristics of the OH stretch vibration of water molecules at the interfaces under investigation. We find an overall blue shift of the OH stretch vibrational band at all interfaces compared to bulk water, indicative of a weakening of the hydrogen bond network at hydrophobic interfaces. We then discuss the vibration of free OH groups, which are known to provide a distinct vibrational signature of water at hydrophobic interfaces across various spectroscopic techniques. We observe a very similar number of free OH groups at all three interfaces, demonstrating the similarity of the interfacial hydrogen bond network. The evaluation of vibrational peak positions of free OH groups reveals unexpected behavior. In the literature, the position of free OH groups is typically correlated to the hydrophobicity of a surface. This is rationalized by the assumption that electrostatic interactions between water and the surface lead to a red-shift of the free OH peak, such that the highest free OH frequency is expected to occur at the air–water interface. Here, we report free OH vibrations at the FSAM–water interface at the same, possibly even slightly higher, frequency than at the air–water interface, challenging this interpretation. Performing time–frequency analysis, we reveal that the vibrational frequency of free OH groups is higher the closer the free OH group approaches the surface fluorine atoms, which indicates that the frequency of free OH vibrations cannot be rationalized in terms of electrostatic interactions alone in the systems under study. Instead other origins of vibrational frequency shifts, like changes in the OH bond potential energy due to strong dispersive interactions or frequency-dependent friction \cite{bruenig2022time} will have to be investigated.

Finally, we demonstrate that the line shape of the free OH spectral peak conveys information about the dynamics of water in the immediate vicinity of the interface. The line width of these features is determined by the lifetime of free OH groups, which, in turn, is similarly limited by the rotational and translational motion of water molecules. Interestingly, the lifetime of free OH groups is considerably longer at the FSAM-water interface than at the air-water and HSAM-water interfaces, resulting in a narrow free OH peak. This lifetime is also correlated with an overall slowing down of water dynamics at the FSAM-water interface. While some degree of dynamical deceleration is typical for hydrophobic interfaces \cite{romero2009effect}, the pronounced slowdown observed in our simulations is rather associated with hydrophilic interfaces, rendering fluorinated SAMs atypical hydrophobic surfaces from a spectroscopic perspective.

\section{Methods}

\paragraph{DFT-MD simulations}

DFT-MD simulations of the SAM-water systems are carried out with the CP2K/7.1 software suite \cite{kuhne2020cp2k} using the BLYP exchange-correlation functional \cite{becke1988density, lee1988development}. The electronic valence density is expanded in the MOLOPT-SR-DZVP basis set \cite{VandeVondele2007}, which is treated in reciprocal space using a plane-wave cutoff of 400~Ry. Atomic cores including core electrons are modeled using Goedecker-Teter-Hutter pseudopotentials \cite{goedecker1996separable}. MD simulations are run using a time step of 0.5~fs for at least 100~ps each, after an additional equilibration time of 10~ps. 

The simulations are conducted in the NpT ensemble, employing the canonical sampling through velocity rescaling algorithm \cite{bussi2007canonical} to maintain the systems at a constant temperature of 300~K. To maintain a constant pressure in the system, we implement the following piston setup: Carbon atoms at the bottom of the SAMs are restrained to a hexagonal grid (Fig.~\ref{fig:fig_1}\,D-E) using harmonic restraint potentials of the form
\begin{align}
    U_\mathrm{res} = \sum\limits_{j>i}^{N_\mathrm{C,res}} \frac{k_\mathrm{res}}{2}(r_{ij}-a)^2 \,
\end{align} 
with the restraint constant $k_\mathrm{res}= 1000~\mathrm{kJ}/\mathrm{mol}\,\mathrm{nm}^2$, and $r_{ij}$ representing the distance between the $i$-th and $j$-th restrained carbon atoms. The lattice constants are chosen to align with the experimentally determined values of a = 4.97~$\mathrm{\AA}$ for HSAMs and 5.9~$\mathrm{\AA}$ for FSAMs \cite{chidsey1990chemical,alves1993atomic, barriet2003fluorinated}. Additionally, a constant force in the $z$-direction is applied to each carbon atom at the bottom of the SAMs, corresponding to a pressure of 1~bar. On the opposite side of the periodic box, a distance-dependent repulsive potential $(z - L_\mathrm{sys})^{-10}$ is exerted on all oxygen atoms of the water molecules, where $L_\mathrm{sys}$ defines the limit of the simulation cell, functioning as a soft wall to the piston. In Sec.~S1 of the supplementary material, we show simulation snapshots that include the periodic simulation cell and discuss further simulation details. 

Orientational distributions of water molecules in Fig.~\ref{fig:orientations} are normalized with respect to their respective homogeneous distribution \cite{sedlmeier2008water}
\begin{align}
    &\frac{1}{2\pi}P_0(\theta_\mathrm{dip}, \theta_\mathrm{OH}) = \\ &\left[\sin\frac{\alpha}{2} -\cos^2  \theta_\mathrm{OH} - \cos^2 \theta_\mathrm{dip} + 2 \cos \theta_\mathrm{OH} \cos \theta_\mathrm{dip} \cos \frac{\alpha}{2}  \right]^{-\frac{1}{2}}\, , \label{eq:homogeneous_dist}
\end{align}
where $\alpha$ is the average HOH angle of a water molecule. 

\paragraph{FF-MD simulations}

FF-MD simulations are conducted with the GROMACS 2024 software suite \cite{abraham2015gromacs}. Electrostatic interactions are treated with the Particle Mesh Ewald method \cite{essmann1995smooth}. Lennard-Jones interactions are truncated beyond an interatomic distance of 1.0~nm. In order to employ a time step of 2~fs, all bonds that include hydrogen atoms are constrained to fixed distances using the SETTLE algorithm \cite{miyamoto1992settle} in the case of water molecules and the LINCS algorithm \cite{hess1997lincs} for hydrocarbon molecules.

In the SAM-water simulations all carbon atoms at the bottom of the SAMs are restrained to a hexagonal grid using harmonic restraint potentials of the form
\begin{align}
    U_\mathrm{res} = \sum\limits_i^{N_\mathrm{C,res}}\frac{k_\mathrm{res}}{2} (\mathbf{r}_i - \mathbf{r}_{i0})
\end{align}
with the restraint constant $k_\mathrm{res}= 1000~\mathrm{kJ}/\mathrm{mol}\,\mathrm{nm}^2$ and $\mathbf{r}_{i0}$ is the reference position of the $i$-th restrained carbon atom.
\paragraph{Extraction of vibrational IR spectra}
As discussed in the main text, anisotropic IR spectra are extracted from polarization densities constructed by binning molecular dipole moments calculated from maximally localized Wannier functions. In practice, we compute the imaginary part of IR spectra using the Wiener-Khintschin theorem on the polarization trajectories $M(t)$:
	\begin{align}
	\chi''_{\alpha \beta}(\omega) = \frac{\omega}{\varepsilon_0 k_\mathrm{B}T L_t} \tilde{M}_\alpha(\omega)\tilde{M}^*_\beta(\omega)
	\end{align} 
with total simulation time $L_t$. 
Real parts are then calculated numerically with the Kramers-Kronig relation according to\, \cite{lucas2012fast}
	\begin{align}
		\chi'_{\alpha\beta}(\omega) = -\mathcal{F}^{-1} \left[i\, \mathrm{sgn}(t) \mathcal{F}[\chi''_{\alpha\beta}(\omega)](t)\right]
	\end{align}
with Fourier-transform $\mathcal{F}[f](\omega) = \tilde{f}(\omega)$. All displayed spectra are smoothed with Gaussian Kernels with a width of up to 10 cm$^{-1}$.

\paragraph{Self-collective decomposition}
Given a set of molecular dipole moments $\boldsymbol{\mu}_i^\mathrm{w}$, the total dipole moment of the water slab can be calculated according to
\begin{align}
    \mathbf{M}^\mathrm{w}_i(t) = \sum\limits^{N_\mathrm{w}}_i \boldsymbol{\mu}_i^\mathrm{w}(t)\, .
\end{align} 
The correlation functions $\Phi(t)$ from Eq.~\ref{eq:pol_corr}, can then be decomposed into a self and a collective part according to
\begin{align}
		\Phi_{\alpha}^\mathrm{self}(t) = \sum\limits_{i=1}^{N_\mathrm{w}} \left\langle \mu_i^{\alpha}(0)\mu_i^{\alpha}(t) \right\rangle \label{eq:self_corr}\, , \\
		\Phi_{\alpha}^\mathrm{coll}(t) = \sum\limits_{i\neq j}^{N_\mathrm{w}} \left\langle \mu_i^{\alpha}(0)\mu_j^{\alpha}(t) \right\rangle\, \label{eq:coll_corr} .
\end{align}

\paragraph{Time-frequency analysis}

Given a trajectory of the length of a free OH bond $d_\mathrm{OH}(t)$, we perform time-frequency analysis by calculating the Gabor transform
\begin{align}
    \tilde{g}_\mathrm{OH}(\tau, \omega) = \int\limits_{-\infty}^\infty \mathrm{d}t\, d_\mathrm{OH}(t) e^{-\frac{(t-\tau)^2}{2\sigma^2} e^{-i\omega t}}
\end{align}
with a window width $\sigma=75$\,fs. We then determine its power spectral density
\begin{align}
    \tilde{G}_\mathrm{OH}(\tau, \omega) = \left\vert \tilde{g}_\mathrm{OH}(\tau, \omega) \right\vert^2\, .
\end{align}
Finally we define the instantaneous frequency of the OH group at time $\tau$, as
\begin{align}
    \nu_\mathrm{inst}(\tau) = \frac{1}{2\pi} \frac{ \int\limits_0^\infty \mathrm{d}\omega\, \omega \tilde{G}_\mathrm{OH}(\tau, \omega)}{\int\limits_0^\infty \mathrm{d}\omega\, \tilde{G}_\mathrm{OH}(\tau, \omega)}\, . \label{eq:instantaneous_freq}
\end{align}
To correlate geometrical observables with the instantaneous frequency, we average them with over a Gaussian Kernel with the same window length $\sigma$. 

The joint probability distributions $P(\nu_\mathrm{inst}, q)$ with geometrical descriptor $q$ in Fig.~\ref{fig:sam:free_OH}\,F-I are locally averaged using
\begin{align}
    \bar{P}(q) = \frac{\int\limits_0^\infty \mathrm{d}\nu_\mathrm{inst}\, \nu_\mathrm{inst} P^2(\nu_\mathrm{inst}, q)}{\int\limits_0^\infty \mathrm{d}\nu_\mathrm{inst}\, P^2(\nu_\mathrm{inst}, q)}\, \label{eq:local_average}\,,
\end{align}
which results in the gray lines in Fig.~\ref{fig:sam:free_OH}\,F-I. Averaging with respect to the squared probability distribution puts a larger emphasis on regions in which the probability distribution takes a high value and therefore eliminates spurious contributions at low frequency, which stem from noisy Gabor time-frequency data.

\paragraph{Gold surface preparation}
A universal single-reflection silicon attenuated total-reflection (ATR) crystal from \textit{IRubis} was covered with a SEIRAS-active gold layer using the chemical deposition method from Miyake et al. \cite{miyake2002electroless}. The silicon surface was first treated with a 1:1:1 (v/v/v) hot solution (90\,$^{\circ}\mathrm{C}$) of \ce{HCl} (37\%, \textit{Carl Roth}), \ce{H2O2} (30\%, \textit{Chemsolute}), and water. The surface was then polished with the help of \ce{Al2O3} powder ($<$50 nm, \textit{Sigma Aldrich}) and treated with a \ce{NH4F} (40\%, \textit{Fluka}) solution for one minute.
Gold deposition was performed by covering the surface with a 1:1:1 (v/v/v) mixture of:

\begin{description}
\item[\textit{A}]
0.3 M \ce{Na2SO3} (\textit{Sigma Aldrich}, 98\%)\\
0.14 M \ce{Na2S2O3} (\textit{Sigma Aldrich}, 98\%)\\
0.1 M \ce{NH4Cl} (\textit{Sigma Aldrich}, 99.5\%)

\item[\textit{B}]
2\% \ce{HF} (\textit{Sigma Aldrich}, 99.9\%)

\item[\textit{C}]
0.03 M \ce{NaAuCl4} (\textit{Sigma Aldrich}, 99\%)
\end{description}

Deposition was performed for 1.5 minutes at 60\,$^{\circ}\mathrm{C}$. To remove impurities from the deposition process, the gold surface was treated electrochemically by performing three oxidation cycles with a scan rate of 0.05 V\,s$^{-1}$ between 1.0 V and 1.4 V versus a \ce{KCl}-saturated \ce{Ag/AgCl} reference electrode using a 0.1 M \ce{H2SO4} solution (\textit{Sigma Aldrich}, ultrapure) as electrolyte.

\paragraph{Experimental spectroscopic characterization}
FTIR spectra were recorded in the range between 850 and 4200 cm$^{-1}$ with a resolution of 2 cm$^{-1}$ using a Vertex 70V spectrometer (\textit{Bruker}) equipped with an MCT detector. The gold surface was first covered with 400 µL of a 100 mM aqueous \ce{Na2CO3} (\textit{Merck}, 99.5\%) solution (pH = 9.5), and a background spectrum was recorded (512 scans). A 400 µL aliquot of a 3 mM solution of the respective thiol compound (vide infra) in the same buffer was then added and thoroughly mixed. The adsorption process was monitored over time by recording a spectrum every 10 s (47 scans) for a total of 2.5 minutes. Characterization was performed for 1-hexanethiol (\textit{Sigma Aldrich}, 95\%) and 1,1,2,2-tetrahydrogen-perfluorohexane-1-thiol (\textit{SynQuest Labs}, 97\%).
vertical
Spectra were baseline corrected using the region between 2360 and 1800 cm$^{-1}$, where no vibrational bands appear. The endpoints of this region were connected to the plateau ends of the spectra at frequencies above the \ce{O-H} stretching region ($>$4060 cm$^{-1}$) and below the monolayer bands ($<$960 cm$^{-1}$) with straight lines. The resulting curve was then smoothed using a moving-window average of 100 points and used as a baseline to correct the experimental data.

\section{Acknowledgments}
This work was funded by the Deutsche Forschungsgemeinschaft (DFG, German Research Foundation)-Project-ID 387284271 - SFB 1349, project C4 to R.R.N. and project C5
to J.H. Computational resources were provided by Hochleistungsrechenzentrum Norddeutschland under Project No. bep00106, as well as the HPC clusters at the physics department and ZEDAT, FU Berlin. 






\bibliography{bibliography.bib}

\end{document}